\documentclass[letterpaper, 11pt]{article} 
\usepackage{hyperref}
\author{Nathaniel Roth, Punit Gandhi, Gloria Lee and Joel Corbo}
\title{The Compass Project: Charting a New Course in Physics Education}
\date{\today}
\begin{document}
\maketitle

It's the middle of August 2011, and a group of incoming students at the University of California, Berkeley, huddle over an audio speaker that is covered in a layer of mysterious, sticky liquid called oobleck. One of the students flips a switch, and suddenly the liquid begins to dance, forming intricate, finger-like extensions that reach up from the speaker's surface before abruptly collapsing back onto it.

The students jump back, surprised by what they see. They haven't even attended their first lecture, but they are already hard at work studying exotic physical phenomena as part of a week-long program sponsored by the \href{http://www.berkeleycompassproject.org}{Compass Project}. 

But what is Compass, and how did that group of students end up studying this liquid-on-a-speaker dance a couple of weeks before the start of their freshman year? To answer that, we have to go back several years.

\section*{In the beginning}

In 2006 three Berkeley physics graduate students\textemdash Angie Little, Hal Haggard, and Badr Albanna---began a series of conversations. All three were troubled by some of the things they noticed in the physics department, including a significant lack of women and minorities among the students and faculty.

``It was amazing that you could count the number of women in most graduate classes on a single hand,'' says Hal. ``With incoming graduate classes of 40 to 50 students, I thought that was absurd. The issue of numbers was even more apparent with underrepresented minorities.''

Badr recalls, ``I remember walking through the hallway in front of the physics department's main office early on in graduate school, and there was a board with the faces of the physics faculty. It immediately struck me that all of the faculty members on the board who were people of color had grown up outside of the United States.''

``I felt my minority status quite sharply,'' says Angie. ``I'm a woman. I'm Native American. Neither of my parents have PhDs. I was on Pell grants because my dad was unemployed when I was an undergrad. When I talked to other physics graduate students, many seemed to have fathers with physics PhDs. Any time I felt like I was struggling academically, it was easy to wonder whether I really fit in.''

In addition to their own experiences, Badr, Angie, and Hal were influenced by the perspectives of the undergraduates in the classes they were teaching. Many students, it seemed, found their introductory physics courses to be demoralizing experiences that drove them away from physics, and often that effect was stronger for women and other traditionally underrepresented students.

In losing those students, Berkeley and the wider physics community lost not only potentially talented individuals but also the diversity of viewpoints useful for confronting challenging, unsolved problems. Even students who succeeded in their first-year classes often did not learn the skills of collaboration, communication, and the creative use of modeling and experimentation. Those skills, necessary to a successful scientist, were not part of the curriculum of traditional introductory physics courses.

``I think I could empathize with what many freshman must be experiencing,'' says Angie. ``I knew that we were losing a lot of incredible students during their transition year from high school to college, particularly those from underrepresented backgrounds, and we all wanted to do something about that.'' The question was, what could they do?

Inspiration struck when Badr and Hal spent several weeks as co-instructors for Berkeley's Pre-Engineering Program (PREP), an intensive program that focuses on preparing incoming engineering students for their introductory calculus, chemistry, and physics courses. The program is academically rigorous, but the part Badr and Hal found most worthy of emulation was its community-building aspect. According to Hal, ``This was what we got really excited about. This was something that we really wanted to bring to the physics department.''

PREP also helped them to see that in order to make the biggest impact in the lives of physics undergraduates, they would have to intervene over the summer between the students' final year of high school and first year of college. Badr explains, ``That moment in your life can be a time when all roads appear open before you, but it doesn't last long. We believed that if we wanted to make a difference for undergraduate students interested in science, then that was the time to do something.''

\section*{A project is born}

Armed with that vision, Angie, Badr, and Hal set out to create a summer program for incoming freshmen interested in the physical sciences. Combining lessons they learned from PREP with a study of the physics education literature, they created a student-driven, active, inclusive learning experience. They harnessed the creativity of teams of graduate students in designing a unique curriculum and learning environment. Most important, they used the power of challenging work and shared experiences to build strong community among the students, both undergraduate and graduate.

The Compass summer program was implemented in August 2007. Under the guidance of a team of graduate student teachers, curriculum designers, resident assistants, and logistics coordinators, 11 incoming freshmen lived and learned together for two weeks before the start of the fall semester. The students worked in groups on a curriculum developed and organized around an open-ended question: ``What can earthquakes tell us about the interior of the Earth?'' Through a series of activities and experiments, they explored wave mechanics, developed a model of a material as a series of masses and springs, and ultimately determined the size of Earth's liquid core using earthquake data provided by the instructors.

The undergraduates were not the only ones working hard that summer. The graduate students discovered that designing the program and curriculum was no easy feat. Nevertheless, they found it valuable, exciting, and instructive. ``Building up the program from scratch was both amazing and incredibly difficult,'' says Angie. ``One of my favorite memories was hiring a teaching staff and going on a retreat together that first summer. We bonded, cooked food together, and talked education together. It was really incredible. Many of those people continued to work with the program for many years after.''

Once the summer program ended, the graduate students felt a clear responsibility to continue to support the students with whom they had just spent such a meaningful two weeks. They developed a suite of services, including a mentoring program that paired each undergraduate with a graduate student mentor, a lecture series in which faculty present their research at a level accessible to undergraduates, office hours during which Compass students can work on and ask questions about their science and math classes, and several social events throughout the year. Many Compass students have remained engaged with the program activities throughout their time at Berkeley. Alongside the program's expansion, the summer programs also continued. Since 2007, six summer programs, with topics as diverse as wind turbines, non-Newtonian fluids, and ``levitating'' Slinkies, have allowed a total of 88 students to begin their college experience as part of the Compass community. In addition to the summer Compass program, a fall course about physical modeling was added in 2009, and a spring course about measurement and error analysis came on board in 2012.

For Badr, the nature of the community formed by the Compass Project is captured in its name: ``Before we hosted the first summer program, I think many people viewed us as some kind of remedial education, but that wasn't what we were about at all. We viewed ourselves as trying to help empower students to find their way forward in college and life by weaving together a community. The name `Compass' really captured that for me without resorting to some complicated acronym. As for `Project' versus `Program,' we debated for a long time. In the end, we settled on `Project' because it communicated that Compass would evolve with the needs and interests of the people involved, both students and teachers.''

Indeed, Compass continues to evolve. For example, this academic year the program piloting an ongoing course for incoming transfer students. Josh Brown and Ilya Esterlis experienced firsthand the difficulties they and other transfer students faced in adjusting to the culture of the Berkeley physics department after their community college experiences. They approached Compass with the idea of creating a course to smooth that transition. The Compass volunteers helped them design, implement, and teach the course and folded the transfer students into Compass's mentoring program and other services. Transfer students, like the rest of Compass's target audience, are more likely to come from underserved communities.

\section*{Mission and impacts}

Compass has stayed true to its core mission of increasing diversity and improving the academic experience of its participants. It has created a strong, supportive community of undergraduate and graduate students who are empowered to shape the organization to meet their needs. Compass also provides opportunities for its participants to develop important scientific and leadership skills that are often absent from the undergraduate or graduate experience.

The numbers tell a convincing story: Of the 88 students who participated in Compass's summer programs from 2007 to 2012, 45\% are female, 19\% are first-generation college students, 26\% are Chicano/Latino, 5\% are African American, and 1\% are Native American; \href{http://www.aip.org/statistics/trends/reports/bachdemograph10.pdf}{for comparison}, only 21\% of physics bachelor's degrees go to women and just 8.3\% to underrepresented minorities. \href{http://www.nsf.gov/statistics/seind02/c2/c2s2.htm#retention}{Additionally}, of freshmen nationwide who enter college with an expressed interested in science and engineering, only 38\% complete a degree in those fields within six years. The Compass Project doesn't yet have six-year completion statistics, but of the 26 students who went through the program and entered Berkeley in 2007 and 2008, 58\% have completed a science or engineering degree. 

As further context, Compass students are not selected based on standardized test scores or their high school GPA, and some of them have not even taken physics in high school. Instead, applicants are selected based on their answers to a series of short essay questions that ask about their academic interests, their experiences with diversity, their excitement and enthusiasm toward science, and their desire to be part of and give back to a community.

Although a causal relationship between Compass and retention cannot yet be definitively established, many members of the Compass community cite the project as a reason that they have stayed in STEM (science, technology, engineering, and mathematics). Ana Aceves, a participant in the 2010 summer program, says, ``Too many times I have doubted my capability to excel in the physical sciences, but time and time again I am told otherwise by my friends and mentors from Compass. They encourage me to overcome obstacles by collaborating with my peers. It is largely because of Compass that I am still an astrophysics major.''

Positive culture, strong community, and the availability of STEM-based extracurricular activities have all been shown to help in the retention of women and minority students in STEM fields (see \href{http://dx.doi.org/10.1353/jhe.0.0019}{Physics Today, September 2003, page 46}). Anecdotal evidence suggests that the project's culture and community are instrumental in providing support that Compass students might otherwise not have gotten.

``Making friends before starting classes at UC Berkeley helped me to adjust to my new life,'' says Harjit Singh, a senior astrophysics and Earth and planetary sciences major. ``In addition, these friends shared most of my math and science classes with me. It seemed natural to form study groups with them while other students worked alone.''

Jenna Pinkham, a junior physics major, said that Compass ``can keep an interest in science alive in the face of soul-crushing introductory physics courses. It's like `Oh, you like science? Well let's go throw Slinkies off a roof just to see what happens.' That curiosity and fun-ness is completely missing from introductory physics courses.''

To better understand the overall experiences of Compass undergraduates, students from the 2010 summer program completed interviews and surveys throughout their first and second years at Berkeley. Parts of the survey directly address students' experiences around culture and community and provide strong evidence that the students feel supported by Compass, both academically and personally. For example, 12 students spoke about community, friendship, or meeting people when asked what was important to them about the project, and 6 students spoke about academic or scientific support. Additionally, 13 students reported seeing each other at least once per week (9 had daily contact), 8 studied together, and 9 lived with another Compass student.

A unique aspect of the Compass community is that it bridges the gap between undergraduate and graduate students. Students at both levels share equally in the group's leadership and decision-making authority. Ana notes, ``I am incredibly fortunate to consider some of the graduate students not only my mentors but also my friends.'' Harjit adds, ``To most undergraduates, graduate students are enigmas who only exist as teaching assistants and nothing else. The fact that Compass is a close-knit community of undergraduate and graduate students is a rarity. I interact with and learn from graduate students as people who are simply older friends.'' Graduate student volunteers find the Compass community just as valuable as the undergraduates do.

Dimitri Dounas-Frazer, a graduate student who taught for the 2011 summer program, explains: ``For me, the biggest perk of being in Compass is that it's a great place to meet other people who are passionate about making science better by promoting equity, diversity, and community\textemdash ideas that are often overlooked in grad school because so much emphasis is placed on doing research and publishing papers. The people I've met in Compass have become some of my closest friends.''

In addition to creating a community, Compass helps develop practical skills. In particular, graduate students are offered a rare opportunity to design and implement curricula for the summer programs and semester-long courses. The experience creates not only better teachers but better scientists: ``Contrary to a common theme in physics graduate schools, which says that teaching can do nothing to advance your career, I believe that the skills you learn from teaching are invaluable: public speaking, clear communication, and collaborative skills are all essential for successful physicists,'' says Hal. And Badr notes that teaching has benefited his research: ``Early on in graduate school, I knew I really enjoyed teaching. With hindsight, I can also see that teaching helped me find the parts of physics that really stoked my passion and guided me to my current research interests.''

Finally, Compass provides its student leadership with many opportunities for professional development. Because the entire organization is run by a team of students, each member must learn some of the skills required to keep Compass going: writing grants, evaluating program impacts, maintaining a web server, reading applications, designing promotional materials, and many more. Those valuable experiences are not part of the trajectory of the typical Berkeley physics student.

Moreover, undergraduate and graduate students share equally in Compass's leadership structure and decision-making, so both groups can benefit from the skill-building process. As Dimitri says, ``I've worked side-by-side with both graduate and undergraduate students on a broad range of projects, including curriculum design and high school outreach activities. I tend to forget that some of my team members are undergraduates because, in these situations, we're all colleagues and peers.''

\section*{The road ahead}

Now in its seventh year, Compass is well established at Berkeley. Summer programs continue to bring in new undergraduates, and new grad students sign up each year to teach and organize. Historically, financial support has come from several sources, some of which change from year to year. Compass leaders are always pursuing fundraising opportunities—and the project gratefully accepts private donations. If the past several months are any indication, however, a large part of the project's focus over the next several years is going to be spreading its philosophy and methodology to other institutions across the country.

In April the project won the American Physical Society's Award for Improving Undergraduate Physics Education. And as participants graduate from Berkeley and move on to other institutions, they are eager to share their ideas. Starting in 2013, Compass graduate students Anna Zaniewski, Dimitri Dounas-Frazer, Josh Shiode, and Joel Corbo will complete their doctorates in physics and start working for the newly created Science Modeling Institute (SMI) in Tempe, Arizona. There, they will create student-driven extracurricular STEM programs at the high school level. In collaboration with the \href{http://modelinginstruction.org}{American Modeling Teachers Association}, they will hold workshops for teachers who want to implement the programs at high schools around the country. In addition, SMI will partner with Compass and ASU to launch a program similar to Compass at other universities.

Academia is changing, and as traditional lecture materials and courses become easily accessible online, universities will need to offer students what they can't get on the web: a strong community of peers who support each other both academically and personally. Whereas any student can watch a YouTube video of an oobleck experiment, no amount of web surfing can replicate the shared experience of a group of students working together to understand that experiment. As Angie put it, ``I've learned how important community is, to have folks around that you can bounce ideas around with, as well as personal ownership, that you create something yourself and get to feel proud of it.''

To find out more about the Compass Project, visit its \href{http://www.berkeleycompassproject.org}{website}. You can also read papers about Compass \href{http://arxiv.org/abs/1203.2682}{here} and \href{http://arxiv.org/abs/1207.6848}{here}. Compass also maintains a \href{http://www.youtube.com/user/berkeleycompass?feature=results_main}{YouTube channel} and a \href{https://twitter.com/compassproject}{Twitter feed}.

\textit{The authors all study physics at the University of California, Berkeley. Nathaniel Roth, Punit Gandhi, and Joel Corbo are graduate students. Gloria Lee is a senior undergraduate physics major.}

\section*{Acknowledgments}
Many people have helped to make the Compass Project such a success. A full list of summer program participants--—students and teachers---is available \href{http://www.berkeleycompassproject.org/people}{here}; many of the listed individuals have contributed to Compass beyond the summer program. Berkeley's physics, astronomy, and Earth and planetary sciences departments have provided funding. Private donors are acknowledged \href{http://www.berkeleycompassproject.org/people/donors}{here}. The authors would also like to thank Ana Aceves, Badr Albanna, Kristin Beck, Dimitri Dounas-Frazer, Hal Haggard, Jesse Livezy, and Angie Little for helping to edit and revise this column.

\end{document}